\documentclass[12pt]{article}
\usepackage{amsmath}
\usepackage[dvips]{graphicx}
\usepackage{epsfig}
\usepackage{amsmath}
\usepackage{amssymb}
\usepackage{graphics,epstopdf}
\usepackage{amsfonts}
\usepackage{amsthm}
\usepackage[usenames]{color}

\definecolor{AV}{rgb}{0.65,0.0,0}
\definecolor{GC}{rgb}{0,0.0,0.65}
\definecolor{WS}{rgb}{0,0.65,0}

\usepackage[
      colorlinks=true,
      linkcolor=blue,
      urlcolor=blue,
      filecolor=blue,
      citecolor=red,
      pdfstartview=FitV,
      pdftitle={},
      pdfauthor={},
      pdfsubject={},
      pdfkeywords={},
      pdfpagemode=None,
      bookmarksopen=true
]{hyperref}

\newcommand{\bm}{\begin{multiline}}
\newcommand{\beq}{\begin{equation}}
\newcommand{\eeq}{\end{equation}}
\newcommand{\beqs}{\begin{eqnarray}}
\newcommand{\eeqs}{\end{eqnarray}}

\newcommand{\ra}{\rightarrow}

\begin{document}

\thispagestyle{empty}

\hfill{}

\hfill{}

\hfill{}

\vspace{32pt}

\begin{center}

\textbf{\Large  Static charged double-black rings in five dimensions }

\vspace{48pt}

\textbf{ Cristian Stelea}\footnote{E-mail: \texttt{cristian.stelea@uaic.ro}}
\textbf{Ciprian Dariescu, }
\textbf{Marina-Aura Dariescu,}\footnote{E-mail: \texttt{marina@uaic.ro}}

\vspace*{0.2cm}

\textit{Faculty of Physics, ``Al. I. Cuza" University}\\[0pt]
\textit{11 Bd. Carol I, Iasi, 700506, Romania}\\[.5em]

\end{center}

\vspace{30pt}

\begin{abstract}

 Using a solution generating technique based on the symmetries of the dimensionally reduced Lagrangian we derive two exact solutions of the Einstein-Maxwell-Dilaton field equations in five dimensions. More specifically, these solutions describe general systems of non-extremally charged static asymptotically flat black bi/di-ring. We compute their conserved charges and investigate some of their properties. As expected, we find that in general there are conical singularities in between the black rings and for the static  solutions considered here they cannot be completely eliminated.
 
\end{abstract}

\vspace{32pt}

\setcounter{footnote}{0}

\newpage

\section{Introduction}

In recent years, higher dimensional black holes have been actively studied. Even though the higher dimensional counterparts of the Schwarzschild spherical black hole and its rotating generalization have been known for some time \cite{Tangherlini:1963bw,Myers:1986un}, notably with Emparan and Reall's discovery of the asymptotically flat black ring solution in five dimensions \cite{Emparan:2001wn} (for a review see \cite{Emparan:2006mm}) one realized that the higher dimensional black holes exhibit a much richer behaviour than their four dimensional counterparts (for reviews see for instance \cite{Emparan:2008eg,Obers:2008pj}). The black ring was the first explicit example of an asymptotically flat black object with non-spherical event horizon. Heuristically, one obtains such a black ring by taking a black string in five dimensions, bending it and connecting its ends to form a circle. A static black ring configuration would normally collapse  to form a black hole with spherical horizon topology; indeed, this is the case in four dimensional asymptotically flat space-times as a consequence of topological censorship \cite{Friedman:1993ty,Galloway:1999bp}. However, in five or more dimensions, the spherical topology of infinity does not constrain that of the black hole horizon \cite{Galloway:1999br}. For instance, in five dimensions geometric considerations restrict the topology to those, such as $S^3$ and $S^2\times S^1$, that admit non-negative scalar curvature \cite{Cai:2001su}. The original black ring system was stabilized against collapse by the centrifugal effects of its rotation along the $S^1$ direction. It belongs to the class of the so-called generalized Weyl metrics \cite{Emparan:2001wk,Harmark:2004rm}, which generally admit $(d-2)$-commuting Killing vectors in $d$ dimensions. The rotating black ring provided the first nontrivial example that known properties of the four-dimensional black holes do not hold in higher dimensions \cite{Emparan:2004wy}. Following the discovery of the rotating black ring, its generalization to black Saturn \cite{Elvang:2007rd} and multi-black rings have been found in five dimensions \cite{Iguchi:2007is,Evslin:2007fv,Elvang:2007hs,Izumi:2007qx}. Concentric supersymmetric black rings in five dimensions were first constructed in \cite{Gauntlett:2004wh,Gauntlett:2004qy}.
 
In higher dimensions, by contrast to the single black hole case, solutions describing general charged multi-black hole systems are scarce. The main reason is that, except in the particular cases where the black holes are extremal/supersymmetric, the known solution generating techniques lead to multi-black hole systems with charges proportional to their masses and, therefore, they cannot describe the most general charged solution for which the masses and charges are independent parameters. Recently, in five dimensions, a solution describing a general double-Reissner-Nordstr\"om solution has been recently constructed in \cite{Chng:2008sr}, generalizing the uncharged solutions given in \cite{Tan:2003jz,Teo:2003ug,Teo:2005wf}. The main purpose of this article is to use the procedure presented in  \cite{Chng:2008sr} to construct in closed form solutions describing generally charged black bi-ring and di-ring in five dimensions.
 
The structure of this paper is organized as follows. We first briefly recall the results of the solution generating technique given in \cite{Chng:2008sr} that will allow us to lift four-dimensional charged static configurations to five dimensional Einstein-Maxwell double-black ring systems. As in \cite{Chng:2008sr}, we shall use the general double Reissner-Nordstr\"om solutions in four dimensions \cite{Manko:2007hi} as a seed solution and lift it to five dimensions. The solution generating method extends easily to the more general case of Einstein-Maxwell-Dilaton (EMD) gravity with arbitrary coupling constant, however, for simplicity reasons, in this work we consider the particular case of Einstein-Maxwell theory. We construct the exact solutions that describe systems of two concentric black rings and discuss some of their properties. Finally, we end with a summary of our work and consider avenues for future research.

\section{The solution generating technique}

Let us recall first the results of the solution generating technique used in \cite{Chng:2008sr}. The main idea of this particular solution generating technique is to map a general static electrically charged axisymmetric solution of Einstein-Maxwell theory in four dimensions to a five-dimensional static electrically charged axisymmetric solution of the Einstein-Maxwell-Dilaton (EMD) theory with arbitrary coupling of the dilaton to the electromagnetic field. One performs first a dimensional reduction of both theories down to three dimensions and, after a careful comparison of the dimensionally-reduced lagrangians and mapping of the scalar fields and electromagnetic potentials, one is able to bypass the actual solving of the field equations by algebraically mapping solutions of one theory to the other. The idea of such a solution generating technique can be traced back to the work of \cite{Galtsov:1995mb} that related stationary axisymmetric vacuum solutions in four dimensions to static charged solutions of the four-dimensional EMD theory with arbitrary dilaton coupling. The novelty of the approach in \cite{Chng:2008sr} was also the use of the scaling symmetry of \cite{Chng:2006gh} to introduce new harmonic factors in the final five-dimensional solution. More precisely, suppose that we are given a static electrically charged solution of the four-dimensional Einstein-Maxwell system:
\begin{eqnarray}  \label{4delectric}
\mathcal{L}_4&=&\sqrt{-g}\left[R-\frac{1}{4}\tilde{F}_{(2)}^2\right],
\end{eqnarray}
where $\tilde{F}_{(2)}=d\tilde{A}_{(1)}$ and the only non-zero component of $\tilde{A}_{(1)}$ is $\tilde{A}_{t}=\Phi$. The solution to the equations of motion derived from (\ref{4delectric}) is assumed to have the following static and axisymmetric form:
\begin{eqnarray}  \label{newinitialmetric1}
ds_{4}^{2} &=&-\tilde{f}dt^{2}+\tilde{f}^{-1}\big[e^{2\tilde{\mu}}(d\rho
^{2}+dz^{2})+\rho ^{2}d\varphi ^{2}\big],  \notag \\
{\tilde{A}_{(1)}} &=&\Phi dt.
\end{eqnarray}

Then the corresponding solution of the Einstein-Maxwell-Dilaton system in five dimensions:
\begin{eqnarray}
\mathcal{L}_{5}=\sqrt{-g}\left[R-\frac{1}{2}(\partial\phi)^2 -\frac{1}{4}%
e^{\alpha\phi}F_{(2)}^2\right],
\label{EMDaction5d}
\end{eqnarray}
where $F_{(2)}=dA_{(1)}$, can be written as:
\begin{eqnarray}  \label{new5Dkfl}
ds_{5}^{2}=-\tilde{f}^{\frac{4}{3\alpha^2+4}}dt^{2}+\tilde{f}^{-\frac{2}{%
3\alpha^2+4}}\bigg[e^{2h}d\chi ^{2}+e^{\frac{6\tilde{\mu}}{3\alpha^2+4}%
+2\gamma-2h}(d\rho ^{2}+dz^{2})+\rho^2e^{-2h}d\varphi ^{2}\bigg],
\end{eqnarray}
while the $1$-form potential and the dilaton are given by:
\begin{eqnarray}
A_{(1)}&=&\sqrt{\frac{3}{3\alpha^2+4}}\Phi dt,~~~~~~~ e^{-\phi}=\tilde{f}^{%
\frac{3\alpha}{3\alpha^2+4}}.
\end{eqnarray}
It can be checked that this solution solves the equations of motion derived from (\ref{EMDaction5d}) as expected, given that (\ref{newinitialmetric1}) is a solution to the equations of motion derived from (\ref{4delectric}). 

Here $h$ is an arbitrary harmonic function; once its form has been specified for a particular solution then $\gamma$ can be obtained by simple quadratures using the equations:
\begin{eqnarray}  \label{gammap1a}
\partial_\rho{\gamma}&=&\rho[(\partial_\rho h)^2-(\partial_z h)^2],~~~~~~~
\partial_z{\gamma}=2\rho(\partial_\rho h)(\partial_z h).
\end{eqnarray}

Solutions of the pure Einstein-Maxwell theory in five dimensions are simply obtained from the above formulae by taking $\alpha=0$. In the following sections, for simplicity, we shall focus on this particular case.

\subsection{The four-dimensional seed solution}

It was shown in \cite{Chng:2008sr} that, using the four-dimensional Reissner-Nordstr\"om as the initial seed, by an appropriate choice of $h$ one can obtain either a black hole, a black ring or a black string in five dimensions.The aim of this work is to construct in closed form the general charged solutions describing static systems of double black rings lying either in the same plane (di-ring) or in orthogonal planes (bi-ring). To this end we shall use the four-dimensional double Reissner-Nordstr\"om solution in the parameterization given recently by Manko in \cite{Manko:2007hi}: 
\begin{equation}
\tilde{f}=\frac{A^{2}-B^{2}+C^{2}}{(A+B)^{2}},~~~~~~e^{2\tilde{\mu}}=\frac{%
A^{2}-B^{2}+C^{2}}{16\sigma _{1}^{2}\sigma _{2}^{2}(\nu
+2k)^{2}r_{1}r_{2}r_{3}r_{4}},~~~~~~~\omega =-\frac{2C}{A+B},  \label{Manko}
\end{equation}%
where:
\begin{eqnarray}
A &=&\sigma _{1}\sigma _{2}[\nu
(r_{1}+r_{2})(r_{3}+r_{4})+4k(r_{1}r_{2}+r_{3}r_{4})]-(\mu ^{2}\nu
-2k^{2})(r_{1}-r_{2})(r_{3}-r_{4}),  \notag \\
B &=&2\sigma _{1}\sigma _{2}[(\nu M_{1}+2kM_{2})(r_{1}+r_{2})+(\nu
M_{2}+2kM_{1})(r_{3}+r_{4})]  \notag \\
&&-2\sigma _{1}[\nu \mu (Q_{2}+\mu )+2k(RM_{2}+\mu Q_{1}-\mu
^{2})](r_{1}-r_{2})  \notag \\
&&-2\sigma _{2}[\nu \mu (Q_{1}-\mu )-2k(RM_{1}-\mu Q_{2}-\mu
^{2})](r_{3}-r_{4}),  \notag \\
C &=&2\sigma _{1}\sigma _{2}\{[\nu (Q_{1}-\mu )+2k(Q_{2}+\mu
)](r_{1}+r_{2})+[\nu (Q_{2}+\mu )+2k(Q_{1}-\mu )](r_{3}+r_{4})\}  \notag \\
&&-2\sigma _{1}[\mu \nu M_{2}+2k(\mu M_{1}+RQ_{2}+\mu R)](r_{1}-r_{2})
\notag \\
&&-2\sigma _{2}[\mu \nu M_{1}+2k(\mu M_{2}-RQ_{1}+\mu R)](r_{3}-r_{4}),
\end{eqnarray}%
with constants:
\begin{eqnarray}
\nu &=&R^{2}-\sigma _{1}^{2}-\sigma _{2}^{2}+2\mu
^{2},~~~~~~~k=M_{1}M_{2}-(Q_{1}-\mu )(Q_{2}+\mu ),  \notag \\
\sigma _{1}^{2} &=&M_{1}^{2}-Q_{1}^{2}+2\mu Q_{1},~~~~~~~\sigma
_{2}^{2}=M_{2}^{2}-Q_{2}^{2}-2\mu Q_{2},~~~~~~~\mu =\frac{%
M_{2}Q_{1}-M_{1}Q_{2}}{M_{1}+M_{2}+R},
\end{eqnarray}%
while $r_{i}=\sqrt{\rho ^{2}+\zeta _{i}^{2}}$, for $i=1..4$, with:
\begin{equation}
\zeta _{1}=z-\frac{R}{2}-\sigma _{2},~~~~~\zeta _{2}=z-\frac{R}{2}+\sigma
_{2},~~~~~\zeta _{3}=z+\frac{R}{2}-\sigma _{1},~~~~~\zeta _{4}=z+\frac{R}{2}%
+\sigma _{1}.
\label{ai}
\end{equation}%
This solution is parameterized by five independent parameters and describes
the superposition of two general Reissner-Nordstr\"{o}m black holes, with
masses $M_{1,2}$, charges $Q_{1,2}$ and $R$ the coordinate distance
separating them. For a detailed discussion of its properties we refer the
reader to \cite{Manko:2007hi} and the references therein. We shall currently note
that, in general, the function $e^{2\tilde{\mu}}$ can be determined up to a
constant and its precise numerical value has been fixed here by allowing the
presence of conical singularities only in the portion in between the black
holes along the $\varphi $ axis. Consequently one has:
\begin{equation}
e^{2\tilde{\mu}}|_{\rho =0}=\left( \frac{\nu -2k}{\nu +2k}\right) ^{2},
\label{strutManko}
\end{equation}%
for $-R/2+\sigma _{1}<z<R/2-\sigma _{2}$ and $e^{2\tilde{\mu}}|_{\rho =0}=1$
elsewhere. This result will be put to use later in describing the presence of the conical singularities in between the black rings in the constructed solutions.

\section{The final EM solution in five dimensions}

Gathering up all the results from the previous section, the corresponding five-dimensional solution of the Einstein-Maxwell system reads:
\begin{eqnarray}  \label{rel1}
ds_{5}^2&=&-\tilde{f}dt^2+\tilde{f}^{-\frac{1}{2}}\bigg[e^{2h}d\chi^2+e^{-2h}%
\big[e^{3\tilde{\mu}/2+2\gamma}(d\rho^2+dz^2)+\rho^2d\varphi^2\big]\bigg],
\notag \\
A_t&=&-\frac{\sqrt{3}C}{A+B}.  \label{final5}
\end{eqnarray}
So far the harmonic function $h$ is still arbitrary and can be chosen at will. One can see that $h$'s presence can alter the rod structure of the final solution along the $\chi$ and $\varphi$ directions. By carefully choosing the form of $h$, one can construct the appropriate rod structures to describe the wanted configurations involving double-black rings. Finally, once one picks a suitable $h$, $\gamma$ is easily found by integrating (\ref{gammap1a}).

\subsection{The di-ring system}

To construct the solution describing a di-ring, \textit{i.e.} a system consisting two concentric black rings, which lie in the same plane, it turns out that one has to pick:
\begin{eqnarray}
e^{2h}&=&(r_0+\zeta_0)\sqrt{\frac{(r_1+\zeta_1)(r_3+\zeta_3)}{(r_2+\zeta_2)(r_4+\zeta_4)}},
\label{diringh}
\end{eqnarray}
where $\zeta_0=z+\frac{R}{2}+\sigma_0$, with $\sigma_0>\sigma_1$. By integrating (\ref{gammap1a}) one finds that $\gamma$ is given by:
\begin{eqnarray}
e^{2\gamma-2h}&=&\frac{2}{K_0r_0}\sqrt{\frac{Y_{20}Y_{40}}{Y_{10}Y_{30}}}
\left(\frac{Y_{12}Y_{14}Y_{23}Y_{34}}{r_1r_2r_3r_4Y_{13}Y_{24}}\right)^{\frac{1}{4}},
\label{diringg}
\end{eqnarray}
where $K_0$ is a constant and we denote $Y_{ij}=r_ir_j+\zeta_i\zeta_j+\rho^2$, where $i,j=0...4$. Replacing these functions in (\ref{final5}) one obtains a general static solution describing a system of two concentric charged black rings lying in the same plane, that is, a general static di-ring system.
\begin{figure}[tbp]
\par
\begin{center}
\includegraphics{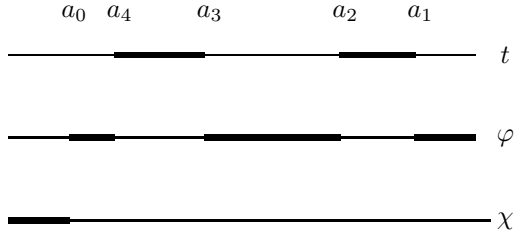} 
\end{center}
\caption{Rod structure of the di-ring system.}
\label{di-rod}
\end{figure}
Before we discuss some of its physical properties, let us first consider the rod structure of this solution. Following the procedure given in \cite{Harmark:2004rm}, one deduces that the rod structure of the general solution is described by five turning points that divide the $z$-axis into six rods as follows.\footnote{We are writing the vectors in the basis $\{\partial/\partial t, \partial/\partial\varphi, \partial/\partial \chi\}$.} 

For $z<-R/2-\sigma_0$ one has a semi-infinite rod with direction $l_1=(0,0,1)$, while for $-R/2-\sigma_0<z<-R/2-\sigma_1$ one has a finite spacelike rod with direction $l_2=(0,1,0)$. For $-R/2-\sigma_1<z<-R/2+\sigma_1$ one has a timelike rod with direction $l_3=(1,0,0)$, which corresponds to the horizon of the first black ring. For $-R/2+\sigma_1<z<R/2-\sigma_2$ on has a finite spacelike rod with direction $l_4=(0,1,0)$. For $R/2-\sigma_2<z<R/2+\sigma_2$ one has a timelike rod with direction $l_5=(1,0,0)$, which corresponds to the horizon of the second black ring. Finally, for $z>R/2+\sigma_2$ one has a semi-infinite spacelike rod with direction $l_6=(0,1,0)$. One should note that the rod directions of the spacelike rods surrounding the black ring horizons are precisely the rod directions corresponding to a system of static black rings whose $S^1$ factor of the horizon is parameterized by $\chi$, while the $S^2$ factor is parameterized by $z$ and $\varphi$. This confirms that the general solution that we derived describes a pair of concentric black rings lying in the same plane, whose rod structure is given in Figure \ref{di-rod}. 

Turning now to the discussion of the conical singularities, one finds that asymptotic flatness fixes the periodicities of the $\chi$ and $\varphi$ coordinates be $2\pi$, while the constant $K_0=4\sqrt{2}$. However, in between the black rings, along the rod direction $l_4$ there is a conical singularity unless:
\begin{equation}
\left( \frac{\nu +2k}{\nu -2k}\right) ^{3}=\left( \frac{(R+\sigma
_{1})^{2}-\sigma _{2}^{2}}{R^2-(\sigma _{2}-\sigma _{1})^{2}}\right)\frac{R+\sigma_0-\sigma_2}{R+\sigma_0+\sigma_2}. 
\label{conditie1-diring}
\end{equation}
Also, the regularity condition along the $l_2$ rod direction imposes the extra condition:
\beqs
\frac{(R+\sigma_0+\sigma_2)(\sigma_1+\sigma_0)}{(R+\sigma_0-\sigma_2)(\sigma_0-\sigma_1)}&=&1.
\eeqs
So far our numerical investigations failed to find values of the physical parameters for which these regularity conditions are satisfied. Physically, the presence of these unavoidable conical singularities is to be expected since our solution is static and, therefore, the conical singularities signal the presence of some other forces needed to balance the gravitational and electromagnetic forces in between the black rings and also to balance the black rings themselves.\footnote{Such conical singularities do appear even the case of a single electrically charged black ring \cite{Kunduri:2004da,Ida:2003wv,Yazadjiev:2005hr,Elvang:2003yy,Yazadjiev:2005gs}.} The only way to satisfy these conditions is to consider extremal rings, for which $k=\mu=0$, $\sigma_1=\sigma_2=0$, however, in this case, the horizons of the black rings become naked curvature singularities.

\subsection{The bi-ring system}

To construct the exact solution describing the charged by-ring system one has to take instead:
\begin{eqnarray}
e^{2h}&=&(r_0+\zeta_0)\sqrt{\frac{(r_1+\zeta_1)(r_4+\zeta_4)}{(r_2+\zeta_2)(r_3+\zeta_3)}},
\label{biringh}
\end{eqnarray}
where $\zeta_0=z-\frac{R}{2}+\sigma_0$ and $\sigma_0<\sigma_2$. By integrating (\ref{gammap1a}) one finds that $\gamma$ is given by:
\begin{eqnarray}
e^{2\gamma-2h}&=&\frac{2}{K_0r_0}\sqrt{\frac{Y_{20}Y_{03}}{Y_{10}Y_{04}}}
\left(\frac{Y_{12}Y_{13}Y_{24}Y_{34}}{r_1r_2r_3r_4Y_{14}Y_{23}}\right)^{\frac{1}{4}},
\label{biringg}
\end{eqnarray}
where $K_0$ is a constant, whose value will be fixed later.

\begin{figure}[tbp]
\par
\begin{center}
\includegraphics{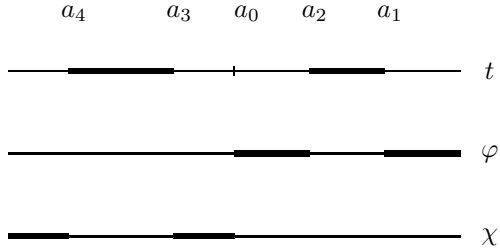}
\end{center}
\caption{Rod structure of the bi-ring system.}
\label{bi-rod}
\end{figure}

As in the di-ring case, one finds that the rod structure of the general bi-ring solution is described by five turning points that divide the $z$-axis into six rods as follows. For $z<-R/2-\sigma_1$ one has a semi-infinite rod with direction $l_1=(0,0,1)$, while for $-R/2-\sigma_1<z<-R/2+\sigma_1$ one has a finite timelike rod with direction $l_2=(1,0,0)$, which corresponds to the horizon of the first black ring. For $-R/2+\sigma_1<z<R/2-\sigma_0$ one has a spacelike rod with direction $l_3=(0,0,1)$. For $R/2-\sigma_0<z<R/2-\sigma_2$ on has a finite spacelike rod with direction $l_4=(0,1,0)$. For $R/2-\sigma_2<z<R/2+\sigma_2$ one has a timelike rod with direction $l_5=(1,0,0)$, which corresponds to the horizon of the second black ring. Finally, for $z>R/2+\sigma_2$ one has a semi-infinite spacelike rod with direction $l_6=(0,1,0)$. This confirms that the general solution that we derived, whose rod structure is given in Figure \ref{bi-rod}, describes a pair of black rings lying in orthogonal planes as expected.

Turning now to the discussion of the regularity conditions along the spacelike rods, one finds that asymptotic flatness imposes the periodicity $2\pi$ for the coordinates $\varphi$ and $\chi$, while the value of the constant $K_0$ is fixed to be $K_0=4\sqrt{2}$. Moreover, to avoid the presence of conical singularities along the rods with directions $l_3$ and $l_4$ one has to satisfy the following two conditions: 
\begin{equation}
\left( \frac{\nu +2k}{\nu -2k}\right) ^{3}=\left( \frac{R^2-(\sigma
_{2}-\sigma _{1})^{2}}{R^2-(\sigma _{1}+\sigma _{2})^{2}}\right)\frac{R-\sigma_0-\sigma_1}{R-\sigma_0+\sigma_1}, 
\label{conditie1-biring}
\end{equation}
respectively,
\beqs
\frac{(R-\sigma_0+\sigma_1)(\sigma_2+\sigma_0)}{(R-\sigma_0-\sigma_1)(\sigma_0-\sigma_2)}&=&1.
\label{condtie2-biring}
\eeqs
Similarly to the di-ring case, we have been unable to find numerically any physically interesting range of the parameters for which the conical singularities disappear. One should note, however, that in the case of extremally charged rings these relations are in fact satisfied. Unfortunately, the horizons of the extremal black rings become again naked curvature singularities and the solutions are pathological.

\section{Some properties of the double-black ring solutions}

For a double-black ring system, the general solution given in (\ref{final5}) has six parameters, which should correspond to the individual masses, electric charges and radii of the constituent black rings. All these quantities can be expressed in terms of the parameters $M_i$, $Q_i$ and $R$ of the four-dimensional seed solution, as well as the new parameter $\sigma_0$ introduced when one picks the explicit form of the harmonic correction $h$.

By construction, our generated solutions are asymptotically flat (as one can also see directly from their rod structure). The asymptotic geometry is found by performing the following coordinate transformations:
\beqs
\rho&=&\frac{r^2}{2}\sin2\theta, ~~~~~z=\frac{r^2}{2}\cos2\theta,
\eeqs
and taking the limit $r\ra\infty$. By using a counterterm prescription or, more directly, by observing the expression of $g_{tt}$ in this limit, one finds that the total ADM mass is given by:
\beqs
M_{ADM}&=&\frac{3\pi}{2G}(M_1+M_2).
\eeqs
The total electric charge of the double-black ring system is evaluated by using the Gauss formula:
\beqs
{\cal Q}=\frac{1}{8\pi G }\int_{S}F_{\mu \nu}dS^{\mu\nu},
 \label{charge}
\eeqs
where $S$ is the three-sphere at infinity, which encloses both the black rings. One finds that ${\cal Q}={\cal Q}^{(1)}+{\cal Q}^{(2)}$, where ${\cal Q}^{(i)}=\frac{\sqrt{3}\pi}{2G}Q_i$ are the individual charges of each black ring, evaluated by integrating (\ref{charge}) over each black ring horizon. One can also evaluate the electric potential $\Phi_H^i=-A_t|_{H}$ for each black ring with the result:
\beqs
\Phi_H^i&=&\sqrt{3}\left(\frac{M_i-\sigma_i}{Q_i}\right).
\eeqs

It is also possible to evaluate the area of each black ring horizon in the general bi/di-ring solutions, by expressing it first in terms of the areas of the black hole horizons of the four-dimensional seed \cite{Manko:2007hi} and making use of the particularly simple form of those quantities, as recently derived in \cite{Manko:2008gb}. More specifically, in the four-dimensional seed solution, the area of each black hole horizon can be expressed in the following form:
\beqs
A_{(4)}^i&=&4\pi\sigma_i\left(\rho\tilde{f}^{-1}e^{\tilde{\mu}}\right)|_{\rho=0}^i,
\eeqs
where for each black hole horizon we have \cite{Manko:2008gb}:
\beqs
\left(\rho\tilde{f}^{-1}e^{\tilde{\mu}}\right)|_{\rho=0}^1&=&\frac{\big[(R+M_1+M_2)(M_1+\sigma_1)-Q_1(Q_1+Q_2)\big]^2}{\sigma_1[(R+\sigma_1)^2-\sigma_2^2]},\nonumber\\
\left(\rho\tilde{f}^{-1}e^{\tilde{\mu}}\right)|_{\rho=0}^2&=&\frac{\big[(R+M_1+M_2)(M_2+\sigma_2)-Q_2(Q_1+Q_2)\big]^2}{\sigma_2[(R+\sigma_2)^2-\sigma_1^2]}.
\label{rhofmu}
\eeqs

For the final five-dimensional solution (\ref{final5}) the area of each black ring horizon can be written as:
\beqs
A_{(5)}^i&=&8\pi^2 \sigma_i\big[(\rho\tilde{f}^{-1}e^{\tilde{\mu}})|_{\rho=0}^i\big]^{\frac{3}{4}}\left((\rho^{\frac{1}{2}}e^{2\gamma-2h})|_{\rho=0}^i\right)^{\frac{1}{2}}.
\label{area}
\eeqs
Let us note now that near each black ring horizon one can expand:
\beqs
e^{2h-2\gamma}&=&(p_i)^2\sqrt{\rho}+{\cal O}(\rho),
\eeqs
where $p_i$ are constants. For the di-ring system one finds explicitly:
\beqs
(p_1)^2&=&\frac{(\sigma_1+\sigma_0)(R+\sigma_2+\sigma_0)}{R-\sigma_2+\sigma_0}\sqrt{\frac{R+\sigma_1-\sigma_2}{\sigma_1(R+\sigma_1+\sigma_2)}},~~~(p_2)^2=\sqrt{\frac{R+\sigma_2-\sigma_1}{\sigma_2(R+\sigma_1+\sigma_2)}},
\label{pi-di}
\eeqs
while for the bi-ring system one finds:
\beqs
(p_1)^2&=&\sqrt{\frac{R+\sigma_1+\sigma_2}{\sigma_1(R+\sigma_1-\sigma_2)}},~~~(p_2)^2=\sqrt{\frac{R+\sigma_1+\sigma_2}{\sigma_2(R+\sigma_2-\sigma_1)}}.
\label{pi-bi}
\eeqs
Then, for the di-ring system, the area of each black ring horizon can be written in the particularly simple form:
\beqs
A_{(5)}^1&=&8\pi^2\sigma_1\bigg[\frac{R-\sigma_2+\sigma_0}{(\sigma_1+\sigma_0)(R+\sigma_2+\sigma_0)}\frac{\big[(R+M_1+M_2)(M_1+\sigma_1)-Q_1(Q_1+Q_2)\big]^3}{\sigma_1(R+\sigma_1+\sigma_2)(R+\sigma_1-\sigma_2)^2}\bigg]^{\frac{1}{2}},\nonumber\\
A_{(5)}^2&=&8\pi^2\sigma_2\bigg[\frac{\big[(R+M_1+M_2)(M_2+\sigma_2)-Q_2(Q_1+Q_2)\big]^3}{\sigma_2(R+\sigma_1+\sigma_2)(R+\sigma_2-\sigma_1)^2}\bigg]^{\frac{1}{2}},\nonumber
\eeqs
while for the bi-ring system one obtains:
\beqs
A_{(5)}^1&=&8\pi^2\sigma_1\bigg[\frac{\big[(R+M_1+M_2)(M_1+\sigma_1)-Q_1(Q_1+Q_2)\big]^3}{\sigma_1(R+\sigma_1-\sigma_2)(R+\sigma_1+\sigma_2)^2}\bigg]^{\frac{1}{2}},\nonumber\\
A_{(5)}^2&=&8\pi^2\sigma_2\bigg[\frac{1}{\sigma_2+\sigma_0}\frac{\big[(R+M_1+M_2)(M_2+\sigma_2)-Q_2(Q_1+Q_2)\big]^3}{\sigma_2(R-\sigma_1+\sigma_2)(R+\sigma_1+\sigma_2)^2}\bigg]^{\frac{1}{2}}.\nonumber
\eeqs
 
To compute the Hawking temperatures of each black ring horizon, one makes use of its definition in terms of the surface gravity evaluated for each individual black ring horizon, $T_H^i=\frac{k_{(5)}^i}{2\pi}$. Here the surface gravity $k_{(5)}$ is generally defined as $k_{(5)}^2=-\frac{1}{2}\xi^{a;b}\xi_{a;b}$, where $\xi=\partial/\partial t$ is the canonically normalized timelike Killing vector. Keeping track of the near horizon expansions of the metric functions $\tilde{f}$ and $e^{\tilde{\mu}}$ in the four-dimensional seed and following the argument given in \cite{Stelea:2009ur} it is easy to show that one can express the surface gravity in the final five-dimensional solution in terms of the surface gravity of the four-dimensional seed as \beqs
k_{(5)}^i&=&p_i(k_{(4)}^i)^\frac{3}{4}\equiv p_i\big[(\rho\tilde{f}^{-1}e^{\tilde{\mu}})|_{\rho=0}^i\big]^{-\frac{3}{4}}.
\eeqs
Then the Hawking temperature of each black ring horizon can be easily calculated in closed form using the expressions listed in (\ref{rhofmu}) respectively (\ref{pi-di}) and (\ref{pi-bi}). For simplicity we shall not list them here.

As it is usually done in multi-black holes systems, one can define the Komar mass of an individual black ring as:
\begin{equation}
M=-\frac{1}{16\pi G }\frac{3}{2}\int_{S}\alpha \ ,
\label{MK}
\end{equation}
where $S$ is the boundary of any spacelike hypersurface and:
\begin{equation}
\alpha _{\mu \nu \rho }=\epsilon _{\mu \nu \rho \sigma \tau }\nabla ^{\sigma}\xi ^{\tau }\ ,
\end{equation}
with the Killing vector $\xi =\partial /\partial t$. In general, this quantity measures the mass contained in $S$, and, therefore, the horizon mass $M_{H}$ is obtained by performing the above integration at the horizon. If one takes instead $S$ to be the three-sphere at infinity enclosing both black ring horizons, then (\ref{MK}) gives the total mass of the system, which coincides with the ADM mass $M_{ADM}$. A straightforward computation leads to:
\begin{equation}
M_{Komar}^{(1)}=\frac{3\pi}{2 G}\sigma _{2},~~~~~~~M_{Komar}^{(2)}=\frac{3\pi}{2 G}\sigma _{1},
\end{equation}
while, following the standard argument,
\begin{equation}
M_{ADM}=M_{Komar}^{(1)}+M_{Komar}^{(2)}-\frac{1}{16\pi G}\frac{3}{2}\int R_{t}^{t}\sqrt{-g}dV.
\end{equation}
However, since Einstein's equations imply $R_{t}^{t}=\frac{F_{\mu t}^{2}}{3}$, one arrives at the following five-dimensional Smarr formula \begin{equation*}
M_{ADM}=\mathcal{M}^{(1)}+\mathcal{M}^{(2)},
\end{equation*}%
where for each constituent one has:\begin{equation*}
\frac{2}{3}\mathcal{M}^{(i)}=T_{H}^{(i)}S^{(i)}+\frac{2}{3}\Phi
_{H}^{(i)}Q_{e}^{(i)}~,
\end{equation*}%
where $\mathcal{M}^{(i)}=\frac{3\pi }{2G}M_{i}$. 

It is now easy to check that this relation follows from the corresponding Smarr formula of the four-dimensional seed solution \cite{Manko:2008gb}, as expected.

\section{Conclusions}

In this paper we made use of the results of a solution generating technique previously derived in \cite{Chng:2008sr} in order to generate, in closed form, the general static solutions describing systems of concentric charged black rings in five dimensions. For simplicity, we focussed on solutions of the Einstein-Maxwell theory only, however, the general solution to the EMD theory with arbitrary dilaton coupling $\alpha$ can be read from (\ref{new5Dkfl}). We also discussed the Weyl rod structures of these solutions confirming that they describe general bi/di-ring systems. We found that the general solutions consisting of static black rings cannot be balanced solely by the effect of their electric charges and, therefore, in absence of rotation there are unavoidable conical singularities in these systems. This is in fact expected to happen on physical grounds, as the effect of the conical singularities is to balance the electromagnetic and gravitational forces in a static system of black objects. We also computed the conserved charges of such systems and proved a Smarr relation, showing that it is in a direct relation to the corresponding Smarr equation in the four-dimensional seed solution.

As avenues for further research, let us note that, similarly to the case of a single black ring system, one expects the static bi/di-ring system to be stabilized and the conical singularities eliminated once one immerses the system in a background electric field. This can be done by first taking the dual of the electric potential to a magnetic $2$-form potential and then applying a Harrison transformation, as described for instance in \cite{Gal'tsov:1998yu}, in order to generate a solution describing a system of two charged black rings in a background magnetic field. Presumably one can then tune the background magnetic field such that the conical singularities are eliminated and the double-black ring system is in equilibrium. The caveat of this procedure is that the double-black rings system is no longer asymptotically flat but instead in a background Melvin-type geometry generated by the background electro/magnetic field. Work on this problem is in progress and it will be reported elsewhere.

Another interesting possibility would be obtaining a  general solution describing a charged double-black ring system in the Taub-NUT background in five dimensions. A simple modification of the technique presented in \cite{Chng:2008sr} has been used in \cite{Stelea:2009ur} to construct double Kaluza-Klein black holes on the double-Taub-NUT background. It would be interesting to find a proper four-dimensional seed solution which, when used in the solution generating technique of \cite{Stelea:2009ur} will provide the black ring metric in the Taub-NUT background. Once that seed is identified, presumably, it can be generalized to a new seed solution that will generate a general double-black ring system in the Taub-NUT background.

\vspace{10pt}

{\Large Acknowledgements}

The work of C. S. was financially supported by POSDRU through the POSDRU/89/1.5/S/49944 contract.

\end{document}